# Understanding Soft Errors in Uncore Components

Hyungmin Cho[1], Chen-Yong Cher[3], Thomas Shepherd[1], Subhasish Mitra[1, 2]
[1]Department of EE and [2]Department of CS
Stanford University, Stanford, CA, USA
[3]IBM T. J. Watson Research Center,
Yorktown Heights, NY, USA

**Abstract**

The effects of soft errors in processor cores have been widely studied. However, little has been published about soft errors in uncore components, such as memory subsystem and I/O controllers, of a System-on-a-Chip (SoC). In this work, we study how soft errors in uncore components affect system-level behaviors. We have created a new mixed-mode simulation platform that combines simulators at two different levels of abstraction, and achieves 20,000× speedup over RTL-only simulation. Using this platform, we present the first study of the system-level impact of soft errors inside various uncore components of a large-scale, multi-core SoC using the industrial-grade, open-source OpenSPARC T2 SoC design. Our results show that soft errors in uncore components can significantly impact system-level reliability. We also demonstrate that uncore soft errors can create major challenges for traditional system-level checkpoint recovery techniques. To overcome such recovery challenges, we present a new replay recovery technique for uncore components belonging to the memory subsystem. For the L2 cache controller and the DRAM controller components of OpenSPARC T2, our new technique reduces the probability that an application run fails to produce correct results due to soft errors by more than 100× with 3.32% and 6.09% chip-level area and power impact, respectively.

**Categories and Subject Descriptors**
B.8.1 Reliability, Testing, and Fault-Tolerance

## 1. Introduction

Radiation-induced soft errors pose a major challenge to building robust systems using complex System-on-Chips (SoCs). Although the soft error rate at the device level (e.g., SRAM cell or latch) stays roughly constant or even decreases over technology generations, the system-level soft error rate increases as more devices are integrated into SoCs [Mitra 14, Seifert 10, 12].

Uncore components[1], such as cache controllers, DRAM controllers and I/O controllers, are increasingly important because their overall area footprint and power consumption in SoCs are comparable to that of processor cores [Gupta 12, Li 13]. The need for studying soft errors in uncore components has been pointed out in the literature [Mukherjee 05, Quinn 13]. While there are many studies on soft errors in processor cores (e.g., [Cho 13, Ramachandran 08, Wang 04]), few have studied soft errors in uncore components. The lack of such studies can be attributed to the difficulties in modeling large-scale SoCs (with multiple processor cores and multiple uncore components) for the following reasons.

1. Uncore studies should model the entire SoC because uncore components interact with processor cores and other uncore components. Modeling only a part of the system may not capture uncore behaviors accurately.
2. Studying system-level effects of soft errors requires real-world applications. This becomes more relevant in the context of cross-layer resilience, where multiple error resilience techniques from various layers of the system stack are combined to achieve cost-effective solutions [DeHon 10, Mitra 10, 14].
3. For statistically significant results, a large number of error injection samples are required. For example, when observing a certain outcome rate, more than 40,000 samples are required to achieve ±0.1% accuracy with 95% confidence when the observed rate is 1%[2].

Such requirements demand high-throughput error simulation or emulation platforms. RTL simulators that model detailed error behaviors are extremely slow. For example, RTL simulation of an out-of-order, superscalar processor core achieves less than a thousand cycles per second [Maniatakos 11b]. High-level simulators, on the other hand, achieve much faster simulation times [Simics]. However, naïvely injecting errors into abstracted high-level layers without adequate low-level details can result in highly inaccurate results (e.g., results in [Cho 13] for processor cores).

Existing uncore error studies are limited to very small designs (e.g., private L1 cache and bus controller in a design with a single processor core [Bailan 10]) or rely on fast high-level simulators without low-level details (e.g., error injections into primary input and output signals in [Graham 09, Lin 06]). While radiation testing can be used to study overall soft error resilience of a design [Bender 08, Sanda 08], it is only available after the chip is produced. Also, quantifying vulnerabilities of various on-chip components can be difficult using radiation testing due to limited observability.

In this paper, we make the following contributions:
1. We present a simulation platform that is capable of simulating large-scale SoCs while modeling detailed flip-flop soft errors[3]. Compared to RTL-only simulation, this platform achieves over 20,000× speedup.
2. We present the first study of system-level effects of soft errors in uncore components in a large-scale OpenSPARC T2 SoC with 500 million transistors, eight processor cores, and many uncore components [OpenSPARC]. We report quantified results on the effects of soft errors in L2 cache controllers, DRAM controllers, crossbar interconnects, and PCI Express I/O controllers. We show that soft errors in uncore components can have significant reliability impact comparable to that of processor cores.
3. We show that traditional system-level checkpoint recovery techniques that generally target processor cores are inadequate for uncore components.
4. We present a new soft error recovery technique called Quick Replay Recovery (QRR). We demonstrate the effectiveness of QRR for the L2 cache controller and the DRAM controller in the OpenSPARC T2 design. QRR results in 100× improvement (i.e., reduction) of the probability that an application run fails to produce correct results due to soft errors; the corresponding chip-level area and power impact for all L2C and MCU instances are 3.32% area and 6.09%, respectively.

The rest of this paper is organized as follows. Section 2 describes our mixed-mode simulation platform and our soft error analysis methodology. Section 3 presents uncore soft error injection results. Section 4 discusses system-level checkpoint recovery challenges for uncore components. Section 5 discusses the accuracy of our mixed-mode simulation platform. Section 6 presents QRR. Section 7 concludes this paper.

## 2. Mixed-mode Soft Error Simulation Platform

To analyze the effects of uncore soft errors in large-scale SoCs, we created a mixed-mode platform that combines two simulation platforms (sometimes referred to as *co-simulation* in design validation literature [Benini 03]). The target uncore component is simulated using an RTL simulator to model soft error behaviors with low-level details, while the rest of the system is simulated using a high-level simulator. Our mixed-mode platform is different from existing co-simulation-based studies on error behaviors for the following reasons:
1. [Li 09, Ejlali 03] use co-simulation to study errors in small combinational logic blocks, such as the ALU or the decoder module with only a few hundred gates, inside a processor core. To correctly model how soft errors in flip-flops behave inside an uncore component, we model an entire uncore component (more than 100K gates) using RTL, and ensure that state transfer between the RTL simulator and the high-level simulator does not become a performance bottleneck.

---
[1] Also be referred to as "nest", "outside-core," or "northbridge". In this paper, we use this term to refer to components that are not processors or accelerators.
[2] This assumes the normal approximation of the binomial distribution, similar to the confidence interval used in [Choi 90].

[3] In this paper, we focus on flip-flop soft errors for the following reasons: a) Design techniques to protect them are generally expensive. Coding techniques are routinely used for protecting on-chip memories. b) Combinational circuits are significantly less susceptible to soft errors [Seifert 12].

2. [Goswami 97, Kalbarczyk 99] profile high-level effects resulting from low-level errors, and use the statistical information for quick error simulations. Profiled error behaviors may not reflect subsequent error propagations due to interactions with the rest of the system (e.g., a flip-flop error in a module may result in multiple erroneous interactions with other components [Cho 13]). We model how the error interacts in a chip by simulating its behavior at the entire chip level until all the effects from the injected error have been fully modeled.

3. [Wang 04] uses two simulators at two different levels of abstraction to simulate a processor core, but only one of the simulators is used at a given point in time. This requires transferring the entire system state between the simulators. In our platform, we utilize low-level simulation only for the target uncore component. This reduces state transfer and low-level simulation overheads.

FPGA emulation platforms can achieve faster speeds compared to RTL simulations while modeling low-level details [Asaad 12, Schelle 10]. However, to model an entire SoC, the design may need to be mapped on multiple FPGA chips. This is because the area required for the FPGA implementation of a design can be an order of magnitude greater than an ASIC implementation (for the same technology generation) [Kuon 07]. As a result, limited inter-FPGA I/O bandwidth can limit the overall emulation speed to only a few MHz [Hauck 10].

## 2.1 Mixed-mode Platform Simulation Modes

Our platform operates in two modes:

1. **Accelerated mode** (Fig. 1a)**:** All components on the chip, including processor cores and uncore components, are simulated using the Simics instruction-set simulator [Simics]. The uncore components are simulated using high-level models. Under error-free conditions, they produce the same output signals to processor cores as the actual uncore components (Fig. 1a ①). Table 1 lists the uncore states modeled by the high-level uncore models (*high-level uncore state*). Flip-flops inside uncore components are not fully modeled in this mode.

2. **Co-simulation mode** (Fig. 1b)**:** The target uncore component is simulated using an RTL simulator. Processor cores access uncore components by exchanging requests and return packets through the on-chip interconnect (e.g., PCX and CPX packets in OpenSPARC T2). During co-simulation mode, these access packets to and from the uncore component are transferred between the high-level simulator and the RTL simulator (Fig. 1b ②). To ensure cycle-level accuracy, the two simulators are synchronized every cycle to ensure transfer of packets between simulators at the correct cycle.

Although the accelerated mode cannot simulate how a soft error behaves at the flip-flop level, high-level models can correctly simulate subsequent behaviors after a flip-flop soft error fully propagates to the high-level uncore state (i.e., no flip-flop or SRAM array inside the uncore component, not included in the high-level uncore state, contains an error).

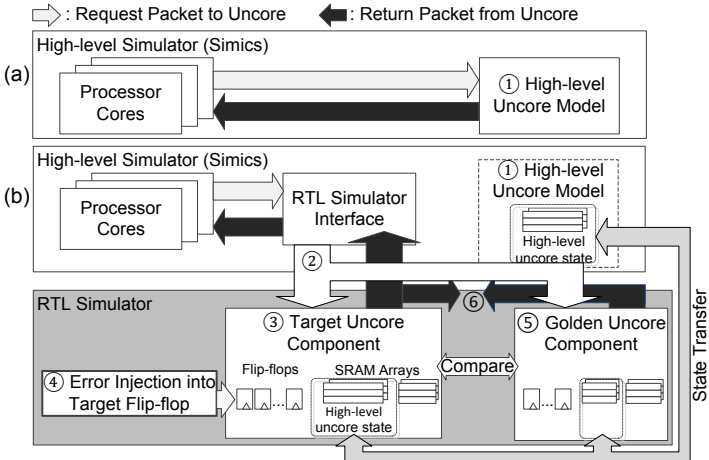

Figure 1. Mixed-mode platforms. (a) Accelerated mode. (b) Co-simulation mode.

Table 1. High-level uncore states modeled by the high-level uncore models.

| Uncore component | High-level uncore states (size per instance) |
|---|---|
| L2 cache controller | Tag address array (28KB), Cache line state bit array (5KB), Cache data array (512KB), L1 cache directory (2KB) |
| DRAM controller | DRAM contents (4GB) |
| Crossbar interconnect | None[4] |
| PCI Express I/O controller | Transfer buffers (RX: 8KB, TX: 4KB) |

## 2.2 Soft Error Injection Methodology

Figure 2 shows the flowchart of our uncore error injection methodology using our *mixed-mode platform*. The co-simulation mode is invoked only when soft error injection begins and terminated when the injected error disappears without any remaining error or when the remaining errors can be simulated using the accelerated mode.

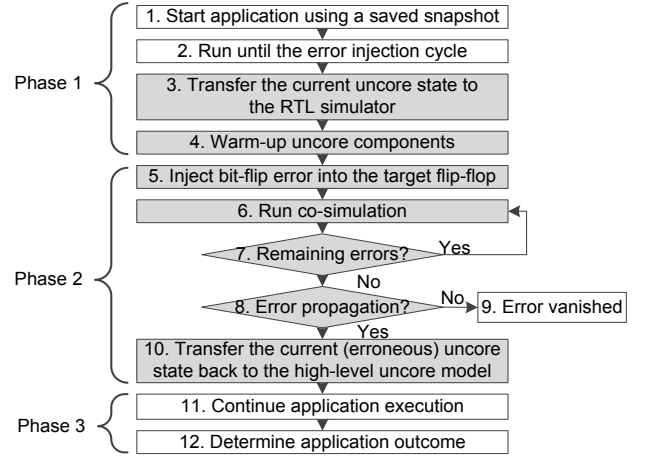

Figure 2. Error injection using our new mixed-mode simulation platform. Steps in grey color uses co-simulation mode.

**Phase 1. Prepare for Error Injection**: For each error injection run, an error injection cycle from high-level simulation (in accelerated mode) and a target flip-flop inside the target uncore component are randomly selected. The mixed-mode platform starts application execution in accelerated mode and simulates the application until the error injection cycle (Fig. 2, steps 1 and 2). This step is shortened by starting the simulation using one of the system state snapshots obtained from a one-time, error-free execution of the application in accelerated mode. If the error injection cycle is $C_i$ and the snapshots are created every $C_f$ cycles, the simulation is started using a snapshot created at cycle $C_s$, where $C_s = \lfloor C_i/C_f \rfloor \times C_f$. For our error injection runs, we created a snapshot every 2 million cycles.

When RTL simulation starts (Fig. 2, step 3), high-level uncore states that have been simulated by the high-level model (Fig. 1a ①) are transferred to the target uncore component in the RTL simulator (Fig. 1b ③). A warm-up period is required before the error injection to correctly restore all microarchitectural states (e.g., flip-flops and small SRAM buffers) that have not been simulated by the high-level model (Fig. 2, step 4). The actual warm-up period is randomly selected for each run to avoid injecting errors always after the same number of co-simulation cycles. In our platform, the warm-up period is at least 1,000 cycles, which is enough to reconstruct microarchitectural states for the tested OpenSPARC T2 uncore components (detailed discussion in Sec. 4.1).

**Phase 2. Inject Error**: A bit-flip error is injected into the selected flip-flop (Fig. 1b ④, Fig. 2, step 5). The platform periodically checks if the accelerated mode can take over the simulation by checking remaining errors in RTL (Fig. 2, steps 6-7). This check is done by comparing the values of the storage elements (flip-flops, SRAM arrays) in the target uncore component, where the error is injected (Fig. 1b ③), with the corresponding values in the golden component (Fig. 1b ⑤). The golden component is an identical copy of the target uncore component that receives the same input, but simulated without error injection. It is only used for simulation purposes to check when to end the co-simulation mode. The co-simulation mode is no longer needed if the comparison finds no mismatch or all mismatches satisfy one of the following conditions:
1. The mismatch can be directly mapped to high-level uncore states. The subsequent effects can be simulated by using the accelerated mode.
2. The mismatch does not cause any functional difference (e.g., corrupted data field when the associated valid flag is not set; the value won't be used by the application in that case).

---
[4] The crossbar interconnect only delivers packets between processor cores and L2 cache controllers. Therefore, its states can be reconstructed in the co-simulation mode without modeling a separate high-level uncore state for the crossbar in the accelerated mode.

**Phase 3. Determine Application Outcome**: The current uncore state in RTL is transferred back to the high-level model, and the mixed-mode platform continues to run the application to completion in the accelerated mode to determine if the application run results in any erroneous outcome (Fig. 2, steps 10-12). Possible outcome types are listed in Sec. 3.2.

During phase 2, the platform monitors if an injected error has produced erroneous return packets to the processor cores by comparing return packets from the target uncore component to those of the golden uncore component (Fig. 1b ⑥). If no erroneous return packet has been detected and the transferred state from the target uncore component matches that from the golden uncore component, the error injection run will result in the same outcome as that of the error-free run. For those cases, the simulation can stop early without executing the rest of the application in phase 3 (Fig. 2, steps 8-9).

## 2.3 Mixed-mode Simulation Performance

The effective simulation throughput of the mixed-mode platform is over 2M cycles/sec, comparable to that of multi-FPGA platforms for large-scale SoCs [Asaad 12, Schelle 10]. Compared to RTL-only simulation of the OpenSPARC T2 design (up to 100 cycles/sec only [Weaver 08]), we achieve more than 20,000× speedup. By utilizing saved snapshots, steps 1-2 take only 1M cycles on average. Steps 11-12 are executed only for less than 1% of total error injection runs[5]. Table 2 summarizes the performance of our mixed-mode platform when simulating an application with cycle length $L$ for the OpenSPARC T2 design. For applications with cycle lengths longer than 280M, the throughput is over 2M cycles/sec. Applications with shorter lengths achieve throughput values less than 2M cycles/sec (e.g., the Radix application with $L$=120M in Sec. 3.2. achieves 1M cycles/sec); however, those applications require shorter simulation times.

$$\text{Throughput} = \frac{\text{Application length}}{\text{avg. simulation time}} = \frac{L}{70 + \frac{L}{4M}} > 2M \text{ cycles/sec}, (L > 280M)$$

Table 2. Mixed-mode simulation performance per each step.

| Simulation type | | Cycles (average) | Performance (cycles/sec.) | Execution time (sec.) |
|---|---|---|---|---|
| Mixed-mode simulation | Steps 1-2 | 1M | 20K | 50 |
| | Steps 3-10 | 10K | 500 | 20 |
| | Steps 11-12 | $L$/2 × 1% | 20K | $L$/4M |
| | Total | | | 70+$L$/4M |

## 3. Soft Error Injection Results for Uncore Components

Using the mixed-mode error injection platform, we performed soft error injection runs for uncore components in the OpenSPARC T2 design (Table 3). In this paper, we study soft errors in the L2 cache controller (L2C), the DRAM controller (MCU), the Crossbar interconnect (CCX), and the PCI Express I/O controller (PCIe)[6].

Table 3. Processor core and uncore components in OpenSPARC T2.

| Component | Number of Instances | Number of Flip-flops (per instance) | Gate count (per instance) |
|---|---|---|---|
| Processor Core | 8 | 44,288 | 513,597 |
| L2C | 8 | 31,675 | 210,540 |
| MCU | 4 | 18,068 | 155,726 |
| CCX | 1 | 41,521 | 370,738 |
| PCIe[7] | 1 | 29,022 | 376,988 |
| NIU | 1 | 135,699 | 1,297,427 |
| SIU | 1 | 16,908 | 105,695 |
| NCU | 1 | 17,338 | 143,374 |

## 3.1 Flip-flops Targeted for Error Injection

Our soft error injection study excludes flip-flops that are already protected or inactive during normal operation. L2C, MCU, and PCIe have built-in error detection and recovery / error correction, such as ECC and CRC, to address errors inside memory arrays. Flip-flops storing ECC or CRC encoded data are effectively protected. Since a single bit-flip in those flip-flops does not affect application-level behavior (after error correction / recovery), they are excluded from error injection. The inactive flip-flops are dedicated to built-in self-test and redundant arrays to repair defective SRAM cells. For this study, we assume a defect-free chip where these flip-flops are not utilized. Table 4 shows the number of flip-flops targeted for error injection in the L2C, MCU, CCX, and PCIe modules.

Table 4. Number of flip-flops in the targeted uncore components.

| Uncore component (number of instances in OpenSPARC T2) | Error injection target flip-flops per instance (% of total flip-flops) | Excluded from error injection | |
|---|---|---|---|
| | | Protected | Inactive |
| L2C (8) | 18,369 (58.0%) | 8,650 (27.3%) | 4,656 (14.7%) |
| MCU (4) | 12,007 (66.4%) | 4,782 (26.5%) | 1,279 (7.1%) |
| CCX (1) | 41,181 (99.2%) | 0 (0%) | 340 (0.8%) |
| PCIe (1) | 23,483 (80.9%) | 5,539 (19.1%) | 0 (0%) |

## 3.2. Benchmark Applications

We use a wide range of multi-threaded benchmark applications: 6 SPLASH-2 benchmarks [Woo 95], 9 PARSEC-2.1 benchmarks[8] [Bienia 11], and 3 Phoenix MapReduce benchmarks for shared-memory systems [Yoo 09] (Table 5). To fully utilize OpenSPARC T2's 64 hardware threads, we instantiated 64 threads for each benchmark application. For PCIe error injections, we modeled a situation where PCIe I/O is used to transfer the application's input data files. In our benchmark set, 12 applications have input data file as shown in Table 5, and they are used for PCIe error injection runs. For each benchmark, we ran more than 40,000 error injection runs for each target uncore component. We assume that only one soft error happens for each application run[9].

Table 5. Benchmark applications.

| | Benchmark application | Error-free execution time (cycles) | Input data file size |
|---|---|---|---|
| SPLASH-2 | Barnes (barn) | 413M | No input file |
| | Cholesky (chol) | 531M | 1.7MB |
| | FFT (fft) | 862M | No input file |
| | LU-contiguous (lu-c) | 215M | No input file |
| | Radix (radi) | 120M | No input file |
| | Raytrace (rayt) | 1,005M | 4.5MB |
| PARSEC-2.1 | Blackscholes (blsc) | 164M | 258KB |
| | Bodytrack (body) | 571M | 2.5MB |
| | Ferret (ferr) | 763M | 4.7MB |
| | Fluidanimate (flui) | 842M | 1.3MB |
| | Freqmine (freq) | 353M | 8.0MB |
| | Streamcluster (stre) | 695M | No input file |
| | Swaptions (swap) | 591M | No input file |
| | Vips (vips) | 1,003M | 7.6MB |
| | X264 (x264) | 881M | 2.8MB |
| Phoenix MapReduce | Linear regression (p-lr) | 54M | 108MB |
| | String match (p-sm) | 248M | 108MB |
| | Word count (p-wc) | 566M | 99MB |

We used the following five outcome categories, used in related studies, to classify application-level outcomes [Cho 13, Sanda 08, Wang 04]: 1) Application Output Not Affected (ONA), 2) Application Output Mismatch (OMM), 3) Unexpected Termination (UT), 4) Hang, and 5) Vanished.

## 3.3. Application-level Erroneous Outcome Rates

Our soft error simulation results demonstrate that uncore soft errors can have significant impact on the overall chip-level soft error rate. Figure 3 shows the observed erroneous outcome rates for each of the uncore components across the benchmark applications and their arithmetic means.

---

[5] Since a run may be terminated or may become unresponsive (UT or Hang outcome type in Sec. 3.2) before step 11, the percentage of runs that require simulation steps 11 and 12 is less than the sum of all erroneous outcome rates presented in Sec. 3.3.

[6] NIU, SIU, and NCU are excluded from this study since RTL simulation of those components requires additional high-level models available only for the Solaris OS on SPARC machines.

[7] Because the OpenSPARC T2 distribution does not provide RTL source of the PCI Express controller, we used an industrial implementation of state-of-the-art PCI Express generation 3 controller design to model soft error effects in I/O controllers.

[8] Facesim application is not tested because the input file for simulation is not included in the benchmark suite. Raytrace application from PARSEC is not tested because it produces no output files, and it is not possible to validate the application results.

[9] The interval between flip-flop soft errors is usually much longer compared to the length of the target benchmark applications [Mukherjee 05]. Actual failure rate of the system can be derived by applying technology-dependent soft error rate to the observed application-level outcome rates per injected soft error.

For example, in Fig. 3a, error injections into L2C for Barnes resulted in 0.42% of ONA, 0.02% of OMM, 1.34% of UT, 0.26% of Hang, and 97.96% of Vanished outcomes.

As expected, most injected soft errors resulted in the Vanished outcome type (over 97% of cases on average). Out of non-Vanished outcomes, UT is the most frequent outcome type for L2C and CCX errors (0.69% on average). However, depending on the application, OMM rates are also significant. For example, the OMM rate for L2C is 0.3% for Fluidanimate and 0.42% for Streamcluster. PCIe error injection results show higher OMM rates (0.89% on average) compared to other components. Since PCIe transfers input data files in our simulations, soft errors in the PCIe likely affect data values. On the other hand, soft errors in other uncore components may corrupt control-related program variables, such as pointers or condition variables that may result in UT or Hang outcomes. Overall, the probability of having an erroneous application outcome (non-Vanished) for a single flip-flop soft error is 1.4%, 1.7%, 2.2%, and 1.7% for L2C, MCU, CCX, and PCIe, respectively.

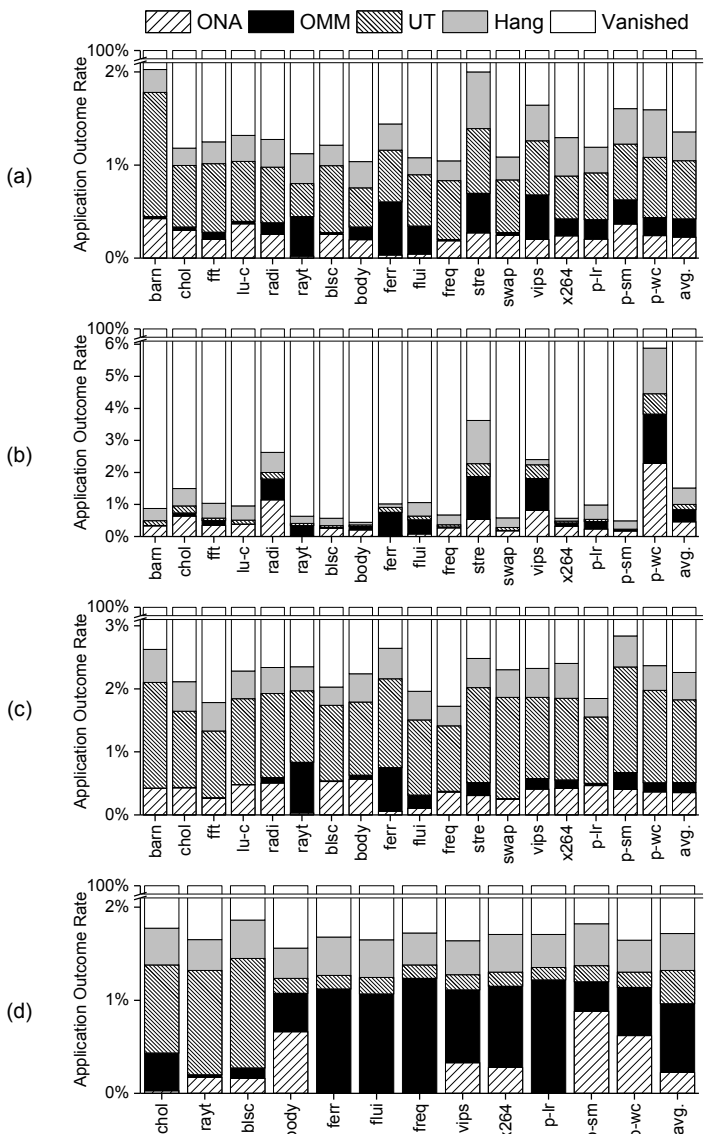

Figure 3. Application-level erroneous outcome rates resulting from error injection for uncore components. (a) L2C. (b) MCU. (c) CCX. (d) PCIe.

The OMM outcome type is a serious reliability concern because, unlike the UT and the Hang outcome types, the user may not be aware that the application resulted in erroneous outputs (unless there are additional mechanisms to verify the correctness of outputs). Figure 4 compares the observed OMM rates obtained from our uncore soft error injection runs to the OMM rates of processor core soft errors reported in the literature[10]. The observed OMM rates of uncore soft errors are comparable to that of processor cores, showing that understanding soft error resilience is important for uncore components in the studied OpenSPARC T2 design.

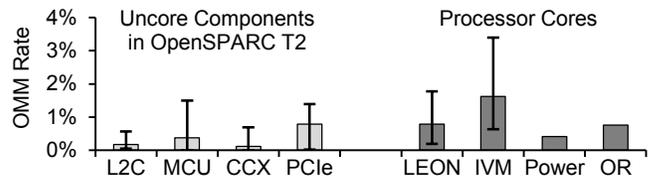

Figure 4. OMM rate of uncore components and processor cores (per instance). Error bars are showing the minimum and maximum values observed across benchmark applications. (LEON: LEON3 SPARC [Cho 13], IVM: IVM ALPHA [Cho 13], Power: IBM POWER6 [Sanda 08], and OR: OpenRISC [Meixner 07]).

## 4. Mixed-mode Platform Accuracy

Unlike RTL-only simulations or FPGA-based emulations, where the system is simulated at the flip-flop level all the time, our mixed-mode platform models detailed flip-flop behaviors only during the co-simulation mode. Hence, it is important to quantify the accuracy of our approach.

### 4.1 Warm-up Period of Co-simulation Mode

To show that only a 1,000 cycle warm-up period is enough to restore the microarchitectural states not included in the high-level uncore model (before an error is injected at the flip-flop), we compared the logic value of each microarchitectural state bit of our mixed-mode simulation setup (during co-simulation mode) vs. a simulation setup that runs the RTL co-simulation from the very beginning (i.e., *full-co-simulation*). In Fig. 5, the Y-axis represents the percentage of bits in our mixed-mode setup (during co-simulation mode) that do not match the corresponding bit in the full-co-simulation mode (unless the bit in the full-co-simulation mode is still unknown). The results are averaged over 10,000 runs. After 1,000 cycles into the co-simulation mode, the microarchitectural state of our mixed-mode platform closely matches that of the full-co-simulation (difference less than 0.2%).

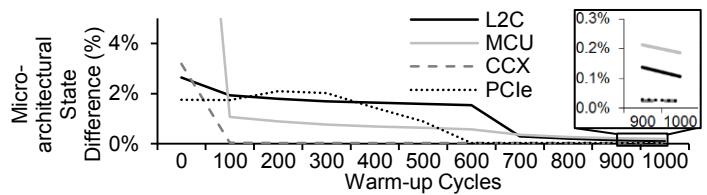

Figure 5. Microarchitectural state difference during the warm-up period.

### 4.2. Limited Co-simulation Length

As discussed before, the co-simulation mode terminates early if the outcome of the application run is determined or if only states modeled by high-level uncore models are erroneous. However, in a few cases, errors may persist in uncore microarchitectural states not modeled by high-level uncore models for extended periods of simulation time. For these cases, limiting co-simulation length is a trade-off between simulation efficiency and accuracy of the obtained results. For our error injection study, only a small subset of soft errors that are injected into a small number of flip-flops result in such situations past 100K cycles of co-simulation. Hence, we limit co-simulation length to 100K cycles. These flip-flops represent 3.7%, 2%, 3.4%, and 3.3% of all flip-flops in L2C, MCU, CCX, and PCIe, respectively (Fig. 6). Out of all error injection runs, only 1.8% actually result in situations in which errors in uncore microarchitectural states not modeled by high-level uncore models persist past 100K co-simulation cycles (L2C: 1.8%, MCU: 0.4%, CCX: 1.5% and PCIe: 1.4% of their respective total runs).

Extending the co-simulation length beyond 100K cycles slows down simulation and has diminishing returns in further determining application outcomes (e.g., extending co-simulation cycle limit by 10× to 1M cycles increases the co-simulation time 10-fold, but the percentage of error injection runs for L2C with errors persisting beyond the cycle limit is reduced from 1.8% to 1.4% only). Since these errors might vanish if given more co-simulation cycles, we do not report them as erroneous outcomes in Figs. 3 and 4. However, one may conservatively choose to protect these flip-flops for error resilient design, as we did in our study of QRR described in Sec. 5.

---

[10] The results are based on injecting one soft error into a single target component (single uncore component or single processor core). The results do not reflect any radiation-hardening techniques or device technologies that have stronger soft error resilience (e.g., SOI [Loveless 11, Oldiges 09]).

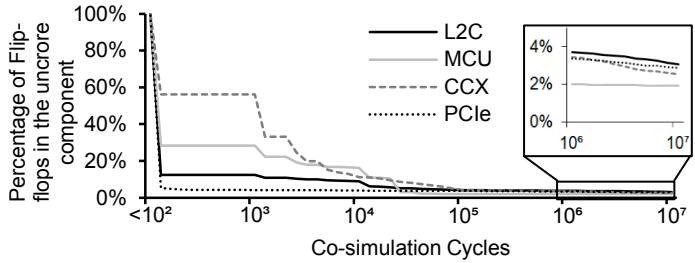

Figure 6. Percentage of flip-flops that result in situations in which errors in uncore microarchitectural states not modeled by high-level uncore models persist beyond the given co-simulation cycles.

### 4.3. Application-level Outcomes Accuracy

We compare the observed outcome rates from our mixed-mode platform vs. those obtained from RTL-only simulations. Due to the slow speed of RTL simulators, the comparison is limited to the FFT application with a smaller data set (1M cycles of execution time), running on 4 threads without an OS. ONA and OMM types are categorized into one outcome type because no specific output generation function (e.g., file write) is implemented in this setup. Figure 7 compares the observed application-level erroneous outcome rates from the two setups obtained from 40,000 error injection samples each. The observed rates from our mixed-mode platform closely matches (0.9-1.1×) those from the RTL-only simulations.

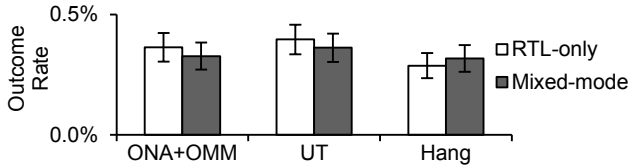

Figure 7. Comparison of observed erroneous outcome rates from RTL-only simulations vs. those from our mixed-mode platform. Error bars represent the 95% confidence intervals.

### 5. System-level Checkpoint Recovery Challenges

Many error resilience solutions depend on system-level checkpoint recovery techniques to revert the system to an error-free state upon error detection [Elnozahy 02]. One major challenge for ensuring correct recovery is the *output commit problem* that may incur long delays for system outputs. Since rollback recovery may not be able to invalidate committed outputs to the outside world, such as network packets or human interactions, outputs should be committed only when it is guaranteed that the system won't roll back to a state before the outputs were produced [Elnozahy 02, Nakano 06]. To avoid such long output delays, two conditions must be satisfied: 1) errors must be detected quickly (short error detection latency) and 2) the recovery operation should not revert the system to a very old state during rollback to an error-free state (i.e., *rollback distance* should be short).

### 5.1. Long Error Detection Latency of Uncore Soft Errors

Error detection techniques at the software and processor architecture levels, such as EDDI [Oh 02] and RMT [Mukherjee 08], can detect uncore errors only after a processor core sees an erroneous output from the uncore component. Therefore, the shortest error detection latency for such techniques is longer than the *error propagation latency to processor cores*, i.e., the duration from the cycle when a soft error affects an uncore component until the cycle when uncore component produces an erroneous output to the processor cores.

For soft errors injected in the uncore components associated with the memory subsystem (L2C, MCU, and CCX) of OpenSPARC T2, we observed very long error propagation latencies (Fig. 5). For example, soft errors in L2C take 36 million cycles to propagate to processor cores on average. For processor cores, in contrast, errors can be detected quickly within a short amount of time [Maniatakos 11a, Smolens 04]. Proactively loading and comparing memory values from uncore components can reduce error propagation and detection latencies, but the associated execution time impact can be high [Lin 14].

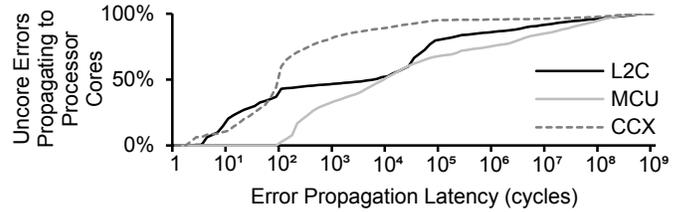

Figure 8. Cumulative distribution of uncore error propagation latencies to processor cores.

### 5.2. Long Rollback Distance for Uncore Soft Errors

To ensure short rollback distance, the checkpointing mechanism has to create checkpoints frequently (short checkpoint interval). To frequently create checkpoints, the data size of each checkpoint has to be kept small due to the limited checkpoint storage size and bandwidth. To achieve small checkpoint data size, *incremental checkpointing* techniques are used [Prvulovic 02, Sorin 02]. They reduce the data size of each checkpoint by saving logs of memory locations [11] modified by processor cores between two checkpoints.

For soft errors in uncore components, however, such techniques may not be adequate. For example, suppose that processor cores modified memory contents in the address range [$X$-$Y$] (and, hence, only those memory contents were included in an incremental checkpoint). However, a soft error in L2C might corrupt the content of memory address $Z$ which is outside the range [$X$-$Y$] (due to an address-related error). In such a case, the recovery mechanism must roll back to an older state with an error-free log on address $Z$.

The required roll back distance to recover from corrupted values in an arbitrary memory location is determined by when a processor core last modified that memory location.

Figure 9 shows the cumulative distribution of required rollback distances resulting from soft errors in L2C and MCU. To cover more than 99% of soft errors resulting in memory corruptions, the required rollback distance can be longer than 400M cycles.

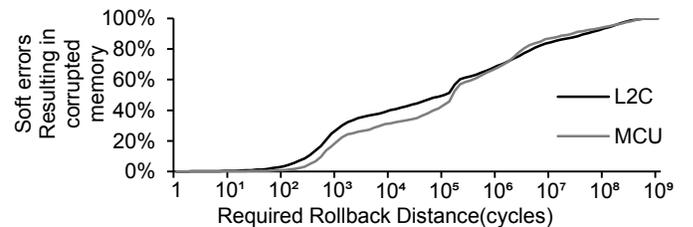

Figure 9. Cumulative distribution of required rollback distance resulting from soft errors in L2C and MCU.

### 6. Uncore Soft Error Resilience Using Quick Replay Recovery

Uncore soft error resilience can be achieved by utilizing radiation-hardened flip-flops [Lilja 13, Mitra 05], but the associated costs can be high (Table 6). Logic parity [Mitra 00] can detect errors with very short error detection latency; combined with an efficient recovery technique, logic parity can provide a low-cost error resilience solution. For processor cores, efficient error recovery techniques exist (e.g., by flushing instructions [Ando 03, Mukherjee 08], or by using instruction-level retry [Meaney 05]). For uncore components, such mechanisms are inadequate due to the following reasons:

1. As discussed in Sec. 2.1, uncore components process request packets from processor cores. Those request packets need to be recreated for recovery. An uncore component may not be able to regenerate request packets by itself.
2. Requesting processor cores to resend request packets may not always be possible since processor cores may not store information about request packets being processed by uncore components. For example,

---
[11] Other architectural states, such as register values, have much smaller size compared to the main memory state, and usually do not require incremental checkpointing.

OpenSPARC T2 processor cores retain request packets only until L2C sends corresponding return packets. However, L2C may continue to process a request even after sending the return packet to the processor core. For example, if a request results in a store miss, L2C may spend hundreds of cycles to fetch a cache line even after sending the return packet. In this case, the uncore operation may be affected by a soft error even after the processor removes the request packet (upon receipt of the return packet).

3. Reverting processor cores, along with the erroneous uncore component, may result in a cascaded rollbacks since each uncore component can interact with multiple processor cores and/or uncore components. For example, rolling back a processor core might require rolling back the uncore components the processor core interacted with, such as other instances of L2C. This, in turn, might require rolling back other processor cores that interacted with those uncore components.

To overcome these challenges, we present a new technique called *Quick Replay Recovery* (*QRR*) targeting uncore components (Fig. 6). QRR handles soft errors without engaging processor cores during recovery. It is applicable for uncore components that satisfy the following properties:

1. Executing requests multiple times in the same order does not change the outcome. For example, this property is maintained in storage components such as memory where duplicated operations in the same order do not change the outcome. (For a detailed discussion regarding this property in the presence of requests accessing the same address, please refer to Sec. 6.3).
2. The uncore component should be able to resume its operation upon reset of its flip-flop contents. For flip-flop contents that should not be reset, such as flip-flops used for configuration bits (e.g., cache disable bit in L2C), radiation-hardening can be selectively used to protect those flip-flops (fewer than 3% for L2C and MCU) from soft errors.

In this paper, QRR works in conjunction with logic parity-based error detection (other error detection techniques with very short error detection latencies are also possible). It provides the following functionality:

1. Record request packets using a record table in the QRR controller. Packets are stored in that table when a new request packet is sent to the uncore component, and deleted from the table when the associated operation is completed by the uncore component (Details in Sec. 6.1). Flip-flops in the QRR controller are protected using radiation hardening.
2. When logic parity detects an error, the QRR controller performs recovery operation by resending the request packets in the record table to the uncore component (Details in Sec. 6.2).

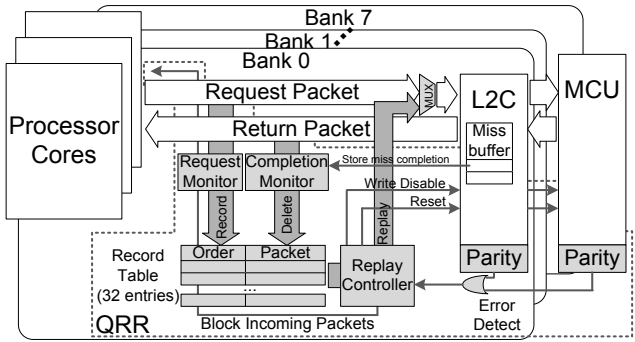

Figure 6. QRR for L2C and MCU. QRR components are shaded.

We evaluate QRR for the L2C and MCU modules for which traditional checkpoint recovery techniques are inadequate (Sec. 4). Because MCU receives access requests through L2C only (e.g., cache line fill, eviction, or non-cached direct DRAM access), recording and replaying L2C requests effectively covers MCU requests as well[12]. QRR incurs a small performance impact during recovery. For L2C, in the worst case when every replayed packet results in the longest operation (L2 cache load miss), the recovery takes fewer than 5,000 cycles.

---
[12] Since an MCU instance operates with two L2C instances in OpenSPARC T2, soft error detection in an MCU invokes recovery operation of two QRR controllers in the two L2C instances.

### 6.1. QRR Normal Operation

During normal operation, the QRR controller keeps track of request packets that are being processed in the uncore component using its record table. QRR for an L2C instance maintains a total ordering of all incomplete requests to that instance based on their arrival order. This is a stricter ordering than the original design, which only needs to maintain the arrival ordering between requests to the same cache line in order to preserve the required SPARC total store ordering (TSO) [OpenSPARC]. Since each L2C and MCU instance exclusively serves disjoint memory address ranges, maintaining ordering at each L2C instance (bank) is sufficient (without affecting requests being processed by other instances).

When requests are completed without errors, they no longer need to be stored by the QRR controller. A completion of a request is determined by monitoring return packets to the processor cores. For uncore requests that require post processing even after the return packet, additional monitoring may be required. In L2C, the only return packet type requiring additional monitoring is a store miss. In this case, the QRR controller waits until the cache miss handling logic (*Miss Buffer*) in L2C completes the operation before deleting the corresponding entry.

### 6.2. QRR Replay Recovery Operation

When logic parity detects an error, QRR first disables write enable signals to data arrays (e.g., L2 cache tag, data, and DRAM) and valid signals of data ports connected to processor cores or other uncore components to prevent the error from corrupting those arrays and propagating to other components.

Propagating the parity error detection signal to the QRR controller and invoking the recovery operation may take multiple cycles because signals from multiple parity detectors have to be aggregated. If a (detected) flip-flop error propagates to a data array or to another component within a few cycles vs. the number of cycles required to propagate the aggregated error signal to the QRR controller, then the soft error might corrupt the corresponding data array or the connected component before the recovery operation is invoked. This creates a non-zero chance of corrupt outputs being produced by the SoC. In our current implementation, we managed this issue by manually inspecting cases where such situations might arise, and fixed the issues by routing individual error signals to disable writes to data arrays and valid signals to other components as needed.

The next step is to assert the reset signal of the uncore component to clear its flip-flop values. Accepting new request packets from processor cores is postponed until recovery is completed. After reset, the QRR controller sends recorded packets to the uncore component in the recorded order until all recorded incomplete request packets are replayed. After the replay completes, the uncore component resumes normal operation by starting to accept new request packets form processor cores.

### 6.3 QRR Correctness

QRR can successfully recover errors for the following reasons:

1. For L2C and MCU, executing incomplete request packets multiple times (replay) does not change the outcome. As long as multiple concurrent requests do not access the same address (i.e., no dependency between concurrent requests), replaying requests in a given order results in the same outcome. If there are dependencies between request packets, L2C is designed not to begin the processing of the following request until the previous one completes (i.e., ordering is maintained).
2. By enforcing a stricter ordering between recorded requests (vs. the default memory ordering of the target uncore component), requests replayed by QRR do not violate the memory access order of the original requests.
3. A detected soft error does not change the outcome of replayed operations since the erroneous flip-flop values are reset by the QRR controller, the contents of the SRAM and DRAM arrays are preserved, and data signals to other components are invalidated (except for the corner case situation discussed in Sec. 6.2).

### 6.4. QRR Results

We implemented QRR for the L2C and MCU modules of OpenSPARC T2, and evaluated its effectiveness using our mixed-mode platform.

To minimize the cost of parity-based error detection, we selectively used radiation hardening for the following flip-flops:

1. Flip-flops with timing slack shorter than the path delay of the XOR tree used to calculate a parity bit. In such a case, logic parity may not be a cost-effective solution since it is not possible to place the XOR tree without slowing down the clock or using additional flip-flops to split the XOR tree over multiple clock cycles. 1,650 flip-flops of L2C (9%), 36 flip-flops of MCU (0.3%) belong to this category.
2. Configuration flip-flops where reset and replay may fail to restore the required flip-flop values. These flip-flops are excluded from reset. 55 flip-flops of L2C (0.3%), 309 flip-flops of MCU (2.5%) belong to this category.
3. Flip-flops in the QRR controller. 812 flip-flops per instance (~3% of the flip-flops in L2C and MCU) belong to this category. Since the flip-flops in the QRR controller are hardened, we did not protect the tables in the QRR controller (assuming single soft errors).

The rest of the flip-flops in the uncore components are protected by logic parity and QRR. After synthesis and place-and-route, the logic area and power overheads [13] of QRR are 45.9% and 47.4% at each uncore component level (3.32% and 6.09% at chip-level for all L2C and MCU instances), which are 23% and 31% lower than the logic area and power costs of protecting all flip-flops using hardening, respectively (Table 6).

Table 6. QRR area and power overhead for L2C and MCU. Flip-flops in the QRR controller are protected using radiation-hardening.

| Overhead | QRR | | | | Hardening only (chip-level) |
|---|---|---|---|---|---|
| | Parity | Hardening | QRR controller and record table | Total (chip-level) | |
| Area | 32.5% | 7.6% | 5.8% | 45.9% (3.32%) | 60.3% (4.34%) |
| Power | 34.8% | 8.7% | 3.9% | 47.4% (6.09%) | 68.3% (8.78%) |

From simulations using the same set of applications as in Sec. 3.2, QRR successfully recovered from all errors injected into the flip-flops covered by logic parity for over 400,000 error injection runs for L2C and MCU[14]. Flip-flops protected using radiation hardening (less than 10% of total flip-flops of L2C and MCU), however, may result in erroneous outcomes since they have non-zero soft error rates. Assuming 1,000× soft error rate reduction of radiation-hardened flip-flops [Lilja 13], the probability of having a flip-flop soft error in the uncore component with QRR is less than 0.013%[15] of that of the same uncore component without QRR. Even with a conservative assumption that all those soft errors result in erroneous (non-Vanished) outcomes, QRR achieves over 100× improvement (i.e., reduction) in the erroneous outcome rate compared to the erroneous outcome rates shown in Sec. 3.3.

## 7. Conclusion

Studying the application-level effects of uncore soft errors in large-scale SoCs is important but difficult. Our new mixed-mode simulation platform enables us to accurately and effectively model uncore soft errors while achieving 20,000-fold speedup compared to RTL simulations. This platform enabled us to characterize, for the first time, system-level effects of soft errors in various uncore components of a large and industrial-grade multi-core SoC.

Our results show that uncore soft errors can have significant impact on the overall reliability of for the studied OpenSPARC T2 multi-core SoC. Hence, resilience techniques to overcome uncore soft errors are required. However, uncore soft errors pose several challenges for traditional system-level checkpointing techniques that are generally effective for processor cores. Our **Q**uick **R**eplay **R**ecovery approach overcomes these challenges for uncore components in the memory subsystem of OpenSPARC T2.

Future research directions include studying system-level effects of a broader class of errors in uncore components (beyond just soft errors), and cross-layer error resilience techniques (spanning circuit, logic, architecture, software, and application layers) for uncore components.

---

[13] The area overhead is obtained using the Synopsys Design Compiler and a commercial 28nm technology library. The power overhead is calculated using the Synopsys PrimeTime and application execution traces. Chip-level overhead is estimated based on published data in related OpenSPARC T2 studies [Jung 14, Li 13].

[14] A more desirable approach is to create a formal proof. With error injection simulations, there can be (rare) corner cases in which QRR may not succeed in recovering correctly from errors.

[15] 90% (flip-flops protected by logic parity detection and QRR recovery) × 0 + 10% (radiation hardened flip-flops) × 1/1,000 + 3% (radiation hardened flip-flops in the QRR controller flip-flops) × 1/1,000= 0.013%

## 8. Acknowledgment

This work was supported in part by Systems on Nanoscale InformatiCs (SONIC), one of the six SRC STARnet Centers, sponsored by MARCO and DARPA. Stanford researchers were supported in part by the National Science Foundation, the Defense Threat Reduction Agency, and the Semiconductor Research Corporation. Stanford and IBM researchers were supported in part by the Defense Advanced Research Projects Agency (Contract No. HR0011-13-C-0022). The views expressed are those of the authors and do not reflect the official policy or position of the Department of Defense or the U.S. Government. This document is Approved for Public Release, Distribution Unlimited.